\documentclass[prl,aps,twocolumn,showpacs]{revtex4}

\usepackage{graphicx}

\begin{document}

\title{Gutzwiller density functional calculations of the electronic
    structure of FeAs-based superconductors: Evidence for a
    three-dimensional Fermi surface}

\author{GuangTao Wang$^{1,2}$, Yuming Qian$^1$, Gang Xu$^1$, Xi Dai$^1$,
Zhong Fang$^1$}

\affiliation{$^1$Beijing National Laboratory for Condensed Matter Physics,
  and Institute of Physics, Chinese Academy of Sciences, Beijing
  100190, China;\\
  $^2$Department of Physics, Henan Normal University, Xinxiang
453007, China }

\date{\today}

\begin{abstract}
  The electronic structures of FeAs-compounds strongly depend on the
  Fe-As bonding, which can not be described successfully by the local
  density approximation (LDA). Treating the multi-orbital fluctuations
  from $ab$-$initio$ by LDA+Gutzwiller method, we are now able to
  predict the correct Fe-As bond-length, and find that Fe-As
  bonding-strength is 30\% weaker, which will explain the observed
  ``soft phonon''. The bands are narrowed by a factor of 2, and the
  $d_{3z^2-r^2}$ orbital is pushed up to cross the Fermi level,
  forming 3-dimensional Fermi surfaces, which suppress the anisotropy
  and the ($\pi,\pi$) nesting. The inter-orbital Hund's coupling $J$
  rather than $U$ plays crucial roles to obtain these results.
\end{abstract}

\maketitle

The iron pnictides are interesting not only because of the high
superconducting transition temperature Tc (above 50K~\cite{1111}), but
also because of their different aspects compared with the high Tc
cuprates: mainly (1) iron pnictides are typical multi-orbital systems,
where spin, orbital and charge degrees of freedom are all
active~\cite{fang}, while in cuprates an effective single band model
can be established; (2) the electron correlation strength is
intermediate~\cite{ARPES-hosono,ARPES,Anisimov}, not as strong as that
in cuprates. However as will be shown in this paper, due to the
multi-orbital nature, where the inter-orbital interaction becomes
important, the correlation in iron pnictides still plays crucial roles
in determining not only the correct internal structure but also the
correct electronic structure near the Fermi Surface (FS).

Both the 1111 (such as LaOFeAs~\cite{1111}) and 122 (like
BaFe$_2$As$_2$~\cite{122}) systems contain the FeAs-layers as the
building blocks, and the low energy bands around the FS are mostly
from the five Fe-$3d$ orbitals~\cite{fang,ARPES-hosono,ARPES,ARPES2}
hybridized with the As-$4p$ orbitals. Since the correlation is not so
strong, the LDA level calculations successfully predict two basic
aspects: (1) the metallic electronic states~\cite{fang}; (2) the FS
nesting and the existence of SDW state in parent
compounds~\cite{SDW,Spm}.  However, LDA calculations encounter
problems with more and more experimental information accumulated. The
most serious problem is that {\it LDA can not describe the Fe-As
bonding accurately (with error bar as large as 10\%)~\cite{Pickett},
but the electronic structures of these compounds depend on the Fe-As
bonding sensitively}~\cite{Geoge}, leaving the calculated electronic
structures so far questionable. Since the possible pairing mechanisms
strongly depend on the detailed electronic structures, the correct
description to the Fe-As bonding and hence the electronic structure is
highly desirable, which is the main subject of our present paper.

The fact that LDA band-width is about two times wider than ARPES
measurement~\cite{ARPES-hosono,ARPES,ARPES2} indicates that the fail
of LDA is mostly due to the insufficient treatment of electron
correlation. Unfortunately, the fully self-consistent treatment of
multi-orbital fluctuation is still a challenging task up to now.  The
correlation effects in iron pnictides have been studied by several
groups~\cite{Anisimov,kotliar} using DMFT (dynamical mean field
theory). Although the structure optimization is not achieved, these
studies indicate that they are in the intermediate coupling region
with the effective mass enhancement varying from $1.5$ to $3.0$. Using
our newly developed LDA+Gutzwiller method~\cite{LDA+G}, where density
functional theory is combined with the Gutzwiller variational approach
with full charge self-consistency, we are able to do structure
optimizations with orbital fluctuations included. It has been
shown~\cite{LDA+G} that, for the ground state determination, the
energy accuracy of Gutzwiller approach is comparable to DMFT. We will
show in this paper that Fe-As bonding can be well described after
considering the correlation: (1) the Fe-As bond length can be
correctly predicted; (2) the Fe-As bonding strength is now estimated
to be 30\% weaker than that of LDA, which explains the experimentally
observed soft phonon modes~\cite{soft-phonon}. We realized that to get
those correct understanding, the Hund's coupling $J$ rather than
Coulomb $U$ is the crucial factor for such multi-orbital systems.

With the correct Fe-As bonding obtained, we study the normal state
electronic structure, and show that FS are quite different with
previous LDA results. The effect of correlation will shift the
$d_{3z^2-r^2}$ orbital up to the Fermi level, as the result
3-dimensional (3D) FS appear in the normal state. The existence of 3D
FS was implied by many experiments, i.e. the low anisotropy in normal
state resistivity, up critical field and superfluity
density~\cite{anisotropy,pen-dep1,anisotropy1,anisotropy2}.  Moreover,
from the calculations of Linhard function, we found that such careful
treatment of orbital fluctuation will reduce the ($\pi,\pi$) FS
nesting effect as well as the effective moment in SDW phase.

The calculations are done with the BSTATE (Beijing Simulation Tool for
Atom Technology) code, in the ultra-soft pseudopotential plane wave
method~\cite{LDA+G}. To guarantee the convergence, we use 30~Ry for
the cut of energy of wave-function expansion, and
12$\times$12$\times$8 K-points mesh in the Brillouin zone. The
projected wannier functions are used as the local orbitals, and the
virtual crystal approximation (VCA) is used for doping.  Since our
local basis is wannier function with all states included, it is
therefore a reasonable value of $U$=3.0-4.0eV, and $J$=0.8-1.0eV, as
suggested by Anisimov~\cite{Anisimov}. Nevertheless, different
parameters has been applied, and optimized values ($U$=3.0eV,
$J$=0.8eV) can be obtained from the correct structure determination.

\begin{figure}[tbp]
\includegraphics[clip,width=8cm]{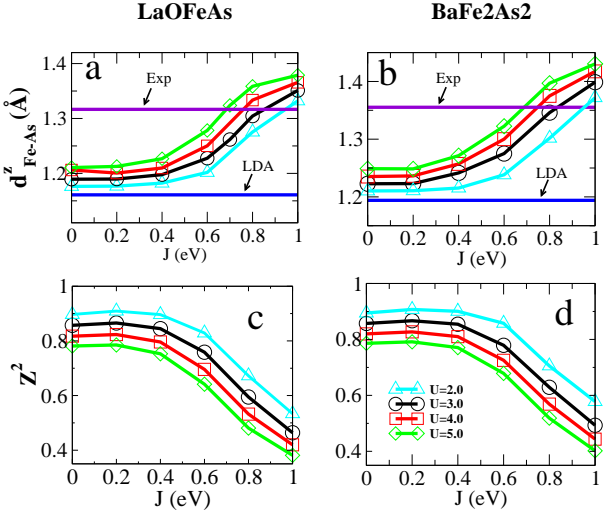}
\caption{The optimized (by LDA+G) Fe-As inter-layer distance
  $d^z_{Fe-As}$ and band-width renormalization factor $Z^2$ as
  function of interactions ($U$ and $J$). The left and the right
  panels are the results for LaOFeAs and BaFe$_2$As$_2$ respectively.}
\end{figure}

{\it 1. Fe-As bond length, phonon frequency, and band narrowing.}

The equilibrium position of As, described by $d^z_{Fe-As}$ (Fe-As
inter-layer distance along the $c$ axis) is studied, and the results
are shown in Fig.1(a) and (b). The LDA $d^z_{Fe-As}$ is 0.15\AA~(for
LaOFeAs) and 0.16\AA~(for BaFe$_2$As$_2$) shorter than their
corresponding experimental value, this is in agreement with previous
calculations~\cite{Mazin-LDA}. However, with the increasing
interaction strength, the $d^z_{Fe-As}$ increase, and approach the
experimental value with the parameters $U\sim$3.0eV and
$J\sim$0.8eV. The same parameters reproduce the As equilibrium
position for both LaOFeAs and BeFe2As2, and the $d^z_{Fe-As}$ is
sensitive to $J$ rather than $U$, implying the important role of
$J$. The same calculations have been repeated for
LaO$_{0.9}$F$_{0.1}$FeAs and Ba$_{0.6}$K$_{0.4}$Fe$_2$As$_2$, and the
same results are obtained.

Not only the Fe-As bond-length but also the Fe-As bonding strength are
seriously affected by the correlation. It has been a puzzling issue
that the measured phonon spectrum of iron pnictides show certain
softening compared with LDA results~\cite{soft-phonon}. There are
three main peaks in the phonon density of states (DOS), located around
13, 23, 30 meV, respectively~\cite{soft-phonon}.  The two peaks (at 13
and 23meV) can be reproduced from LDA, however the peak around 30meV,
which are mostly due to Fe-As bond related modes, is 15\% softer than
LDA results. This ``soft phonon'' energy can be obtained after
reducing the Fe-As bonding strength (force constants) by 30\% from its
LDA values artificially. Here we show that this weakening is naturally
explained by the correlation effect. Fig.2(a) shows the calculated
total energy as function of As-displacement. The curves can be nicely
fitted with parabolic equations, suggesting the minor anharmonical
effect. From the second derivative of the curves, we can estimate the
force constant, which is 30\% smaller in LDA+G comparing with
LDA. This reduced force constant will exactly reproduce the observed
``soft phonon'' around 30meV. We have also optimized the La-position
in the same way, and found that its equilibrium position and force
constant are little affected by the LDA+G treatment.

\begin{figure}[tbp]
\includegraphics[clip,width=7.5cm]{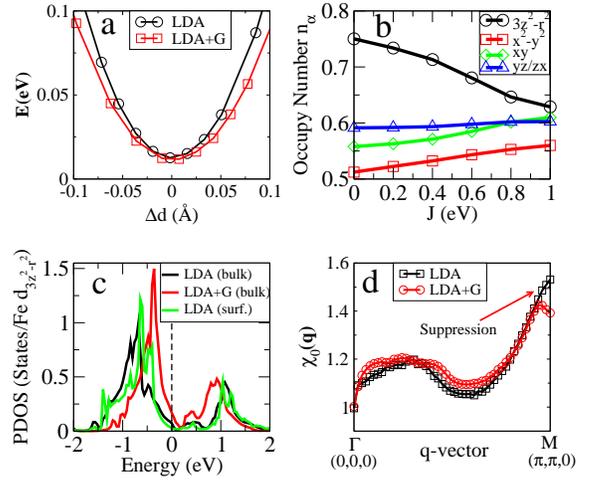}
\caption{(a) The total energy versus As-displacement ($\Delta d$) from
  its equilibrium position. (b) The occupation numbers of five Fe-$3d$
  orbitals as function of $J$. $U$ is fixed to be 3.0eV. (c) The
  projected density of states (PDOS) of Fe-$3d_{3z^2-r^2}$ orbital for
  different situations: LDA for bulk, LDA+G for bulk, and LDA for
  surface. Please note the relative change of $3d_{3z^2-r^2}$ weight
  at the Fermi level. (d) The Lindhard response functionn $\chi_0(q)$
  for $q$-vector along the $\Gamma-M$ line.}
  \label{fig2}
\end{figure}

It has been argued in literature that the discrepancies discussed
above can be improved if spin-polarization is assumed~\cite{Yil-spin},
while our strategy is to treat the system without such
assumption. This is of course the true experimental situation, where
the normal state above transition temperature is paramagnetic
metal. In the presence of orbital fluctuation, the essence of
Gutzwiller approach is to suppress the weight of double-occupancy, and
renormalize the kinetic energy by the factor $Z^2$ (averaged through
five orbitals) as shown in Fig.1(c) and (d). For the $U$ and $J$
determined from structure optimization, the renormalization is about
0.6, which is consistent with the ARPES
results~\cite{ARPES-hosono,ARPES,ARPES2}. (The $Z^2$ factor is again
not sensitive to $U$, but sensitive to $J$). The narrowed band-width
will favor the weaker Fe-As bonding and longer bond-length. It is
worth to note that assuming spin polarization is in fact the simplest
way to suppress the double-occupancy although it is no longer
realistic for the normal state of superconducting compounds.

\begin{figure}[tbp]
\includegraphics[clip,width=8cm]{Fig3a.eps} \\
\includegraphics[clip,width=3.5cm]{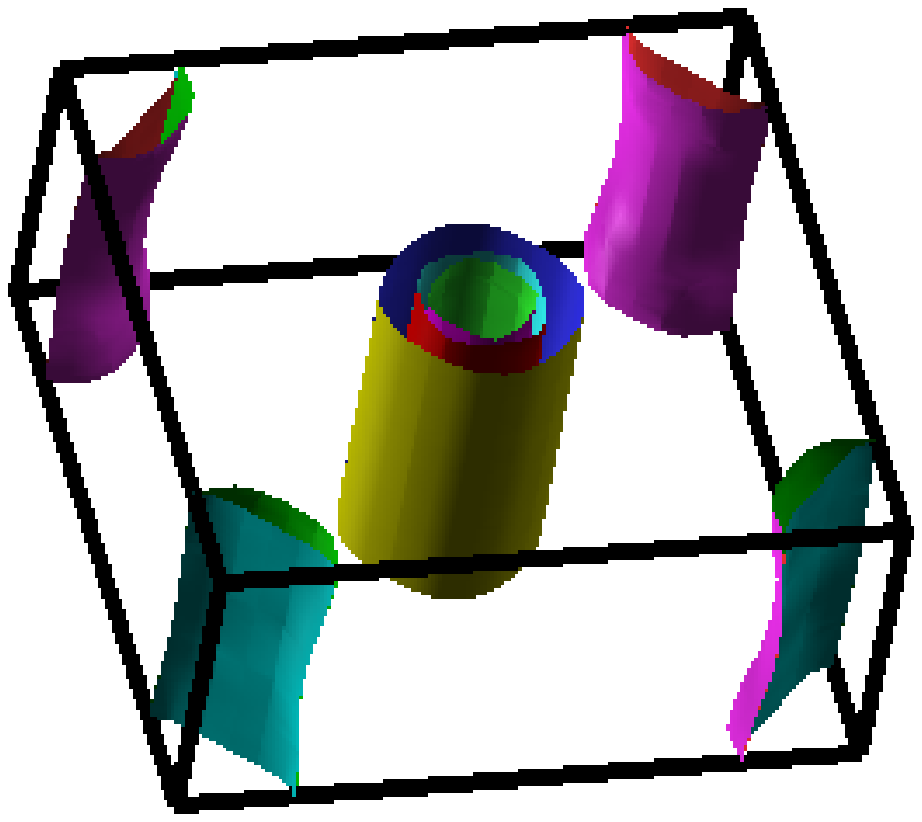}
\includegraphics[clip,width=3.5cm]{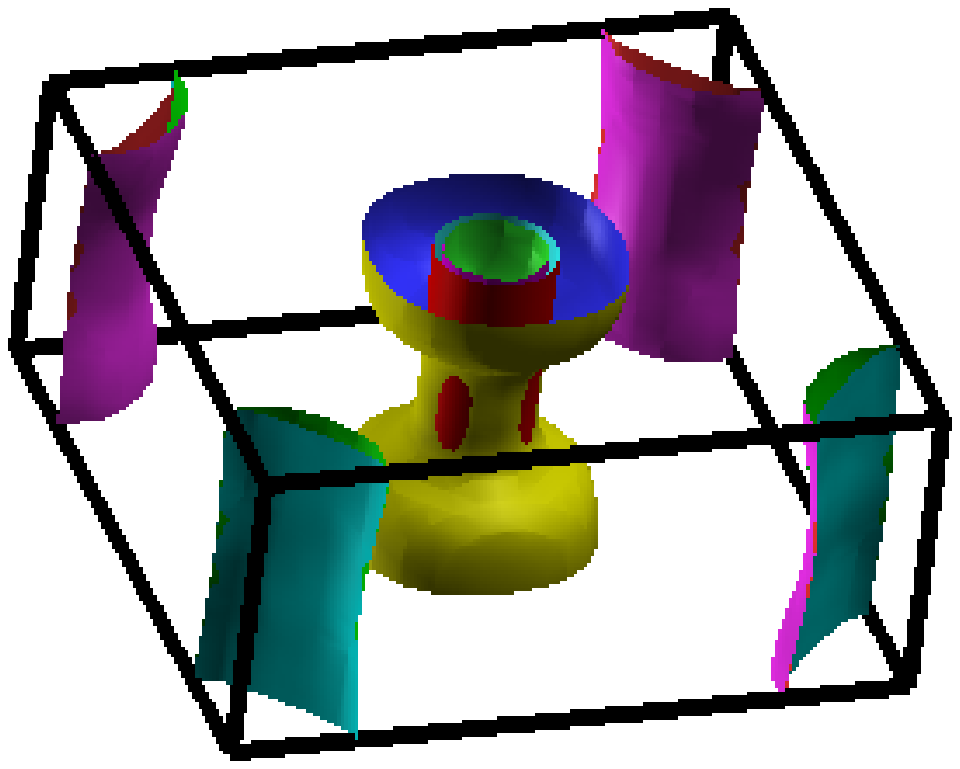}  \\
LDA  \ \ \ \ \ \ \ \ \ \ \ \ \ \ \ \ \ \ \ \ \ \ \ \   LDA+G
\caption{The electronic structure of LaO$_{0.9}$F$_{0.1}$FeAs. Upper
  and middle panels: Band Structures by LDA and LDA+G respectively. We
  plot the fat bands, where the symbol size corresponds to the
  projected weight of Bloch states onto the Fe-$3d_{3z^2-r^2}$
  orbital. Lower panels: Fermi surfaces from LDA (left) and LDA+G
  (right). The $\Gamma$ point located at the body center of the
  cubes.}
\end{figure}

{\it 2. The role of $d_{3z^2-r^2}$ orbital.}

It is essential to understand the role of Hund's coupling $J$ and
discuss the change of orbital characters in the presence of
interactions. Starting from the structure aspect, the crystal
splitting of $d$ orbitals is not so strong (of the order 0.1-0.2eV),
nevertheless the squeezing of FeAs$_4$ tetrahedral along $c$-axis push
the $d_{3z^2-r^2}$ orbital down to the lowest energy among five, and
it contribute little to the DOS at Fermi level ($E_f$). In most of the
effective models studied so far, this orbital is neglected for
simplicity. In the presence of interactions, the role of Coulomb $U$
is to enhance the orbital polarization, however the role of
inter-orbital Hund's coupling $J$ is opposite: it try to tight all
orbitals together and favors even distribution of the $6$ electrons
among five orbitals.  As shown in Fig.2(b) and (c), with the
increasing of $J$, the energy level of $d_{3z^2-r^2}$ is raised and
its occupation is reduced. For the optimal $J$=0.8eV, all orbitals are
almost equally occupied.  The reduced orbital polarization will
enhance the inter-orbital fluctuation for such multi-orbital
system. This is the key to understand the kinetic energy
renormalization factor $Z^2$ around 0.6.

The $d_{3z^2-r^2}$ orbital is now relevant to the low energy physics,
very different with previous understanding based on LDA. Most
significantly, 3D FS appears and the ($\pi,\pi$) FS nesting is
suppressed. The calculated Linhard response funciton shown in Fig.2(d)
suggest that the nesting is not as strong as in LDA, and slightly
incommensurate. The SDW solution with smaller magnetic moment
(comparing with LDA) would thus be expected for the parent compounds.

\begin{figure}[tbp]
\includegraphics[clip,width=8cm]{Fig4a.eps} \\
\includegraphics[clip,width=3.5cm]{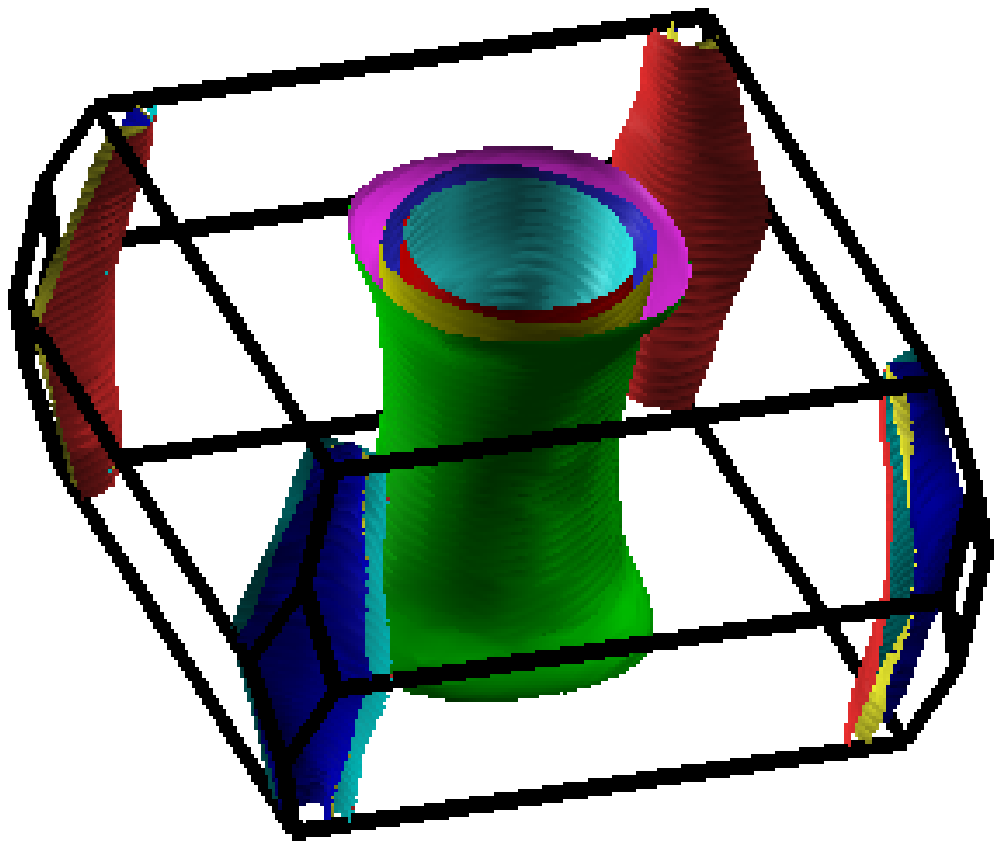}
\includegraphics[clip,width=3.5cm]{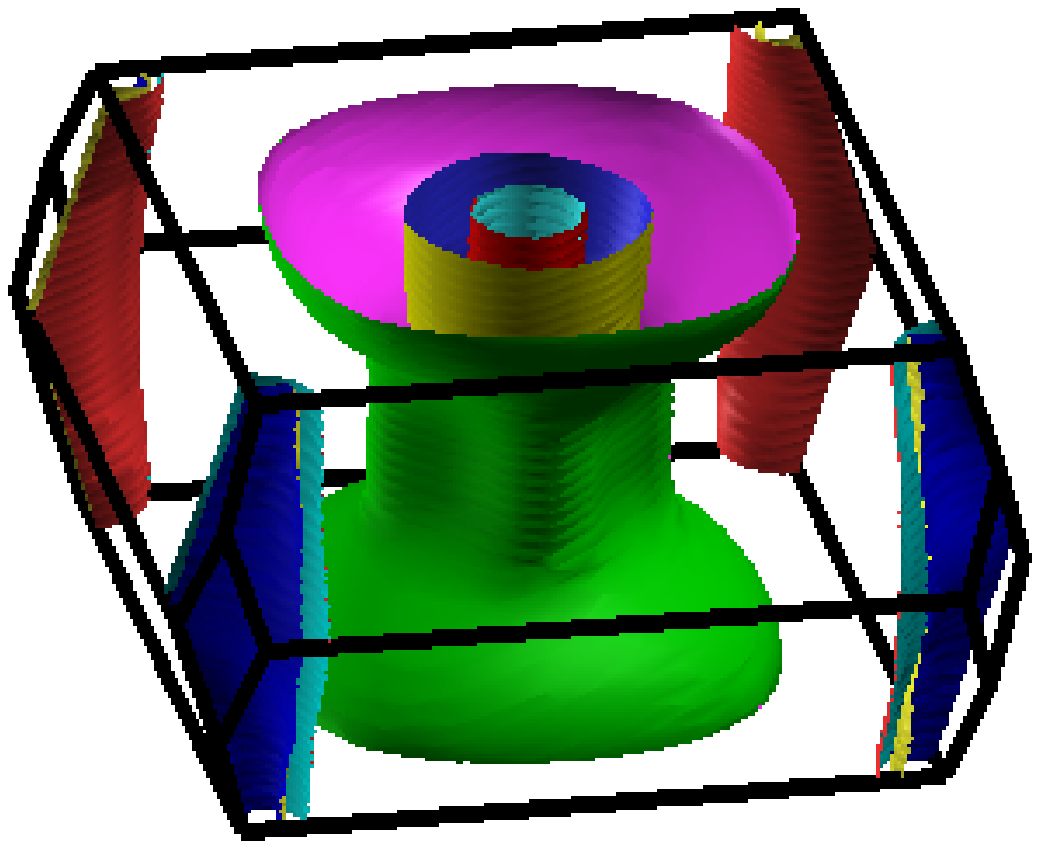}   \\
LDA  \ \ \ \ \ \ \ \ \ \ \ \ \ \ \ \ \ \ \ \ \ \ \ \   LDA+G
\caption{The electronic structure of
  Ba$_{0.6}$K$_{0.4}$Fe$_2$As$_2$. All notations are the same as
  Fig.3. For the comparison reason, we plot the band structures in the
  same high symmetry lines as that in LaOFeAs, instead of the BZ of
  the body centered structure.}
\label{fig4}
\end{figure}

{\it 3. The band structures and FS.}

The normal state band structures of superconducting compounds,
LaO$_{0.9}$F$_{0.1}$FeAs and Ba$_{0.6}$K$_{0.4}$Fe$_2$As$_2$, are
shown in Fig. 3 and 4. The calculations are all done with the
experimental As-position, rather than the incorrect position optimized
from LDA. In the LDA results, the bands which has strong
$d_{3z^2-r^2}$ component are away from the $E_f$, and the FS are
highly 2-dimensional. In the LDA+G results, however, those bands are
clearly shifted up and go cross the Fermi level. As the results, one
of the FS is strongly 3D due to the dominate $d_{3z^2-r^2}$
character. The enhanced $d_{3z^2-r^2}$ contribution to the DOS at
$E_f$ can be quantitatively obtained from Table.I.

\begin{table}
  \caption{Some calculated parameters: 
    $\frac{\sigma_{xx}}{\sigma_{zz}}$ is the conductivity ratio between 
    $a$ and $c$, $N_{3z^2-r^2}$ is the partial DOS of $d_{3z^2-r^2}$ orbital, 
    and $N_{total}$ is the total DOS, at the Fermi level. }
\begin{tabular}{c|c|c|c|c|c|c}
\hline
    &\multicolumn{2}{|c|}{LaO$_{0.9}$F$_{0.1}$FeAs}  
    &\multicolumn{2}{|c|}{Ba$_{0.6}$K$_{0.4}$Fe$_2$As$_2$}
    &\multicolumn{2}{|c}{BaFe$_{1.84}$Co$_{0.16}$As$_2$}   \\ \hline
    &LDA   &LDA+G   &LDA    &LDA+G   &LDA  &LDA+G    \\  \hline
$\frac{\sigma_{xx}}{\sigma_{zz}}$ 
    &45.9  &15.8   &11.5   &2.9    &8.4    &2.7     \\ \hline
$\frac{N_{3z^2-r^2}}{N_{total}}$  
    &2\%   &23\%   &12\%   &22\%   &8.3\%  &14.5\%  \\ \hline
\end{tabular}
\label{tab1}
\end{table}

{\it 4. Comparison with experiments.}

Although most of the iron pnictides have layered structure, the
anisotropy reflected in transport and superconducting properties is
quite low. For instance, in Ba(Fe$_{0.926}$Co$_{0.074}$)$_2$As$_2$,
the anisotropy in resistivity is around 3$\pm$1~\cite{anisotropy}, the
anisotropy in penetration depth is around 3 to
6~\cite{pen-dep1,anisotropy1}, and that of the up critical field for
hole doped compound Ba$_{1-x}$K$_{x}$Fe$_2$As$_2$ is around 2 to 3
just below the critical temperature but down to unity around $10K$
\cite{anisotropy2}.  The inelastic neutron scattering measurement also
show strong 3D features in the resonance peaks appearing in the
superconducting phase~\cite{Neutron}.  Those low anisotropy results
can be naturally explained by the existence of the 3D FS, which
appears only when the correlation effect has been carefully included
with the correct experimental structure. We have calculated the
resistivity anisotropy based on the band structure obtained by LDA+G
assuming the isotropic relaxation time. The results summarized in the
Table I fit quite well with the experimental results. In addition, the
penetration depth\cite{pen-dep1}, $1\over T_{1}T$\cite{NMR1,MuSR}, and
thermal conductivity measurements\cite{Thermal} imply there may be
line nodes in the superconducting states. From our results, we propose
that, at least for the Ba$_{0.6}$K$_{0.4}$Fe$_2$As$_2$, the 3D FS at
$k_z=\pi$ plane is big enough to cross the node lines of the $S_{\pm}$
pairing state\cite{Spm}. Therefore the line nodes may exist on this
hourglass like 3D FS in the $S_{\pm}$ pairing state.

On the other hand, however, the ARPES results~\cite{ARPES,ARPES2}
suggest discrepancies with transport measurements: (1) all the
detected four FS are cylinder like with very weak dispersion along the
$c$-axis; (2) the superconducting gaps in all the four FS are
isotropic without any line nodes. It is still too early to address
these discrepancies, while one possible scenario based on our
calculations is the following.  Since the main component of the
hourglass like 3D FS is $d_{3z^2-r^2}$, which disperses strongly along
the $c$-axis, it should be strongly affected by the surface. To see
the general tendency of surface effect, we have done the surface
calculation using LDA only. We can see from Fig.2(c), the PDOS of
$d_{3z^2-r^2}$ at $E_f$ is strongly suppressed on the surface, which
makes it very difficult to be detected directly by
ARPES. Nevertheless, the possible $k_z$ dependence of FS has been
partly observed in the recent ARPES measurement~\cite{ARPES3}.

In summary, by properly treating the electron correlation through
self-consistent LDA+Gutzwiller method, we are now able to describe the
Fe-As bonding (both bond-length and bonding-strength)
successfully. Having this prerequisite, the resulting electronic
structure is very different with previous LDA understanding. In
particular, the bands are narrowed by a factor of 2, and the
$d_{3z^2-r^2}$ orbital is now relevant to the low energy physics by
contributing to a 3D FS, which suppress the anisotropy and the
($\pi,\pi$) FS nesting. Although the interaction strength in iron
pnictides is not as strong as in cuprates, the inter-orbital Hund's
coupling $J$ (due to the multi-orbital nature) plays crucial roles to
determine the electronic structure.

We acknowledge the supports from the NSF of China, the National Basic
Research Program of China, and the International Science and
Technology Cooperation Program of China.

\end{document}